\documentclass[preprint]{aastex}

\font\ti=cmti10 scaled \magstep1

\received{{\ti 4-04-01}}
\accepted{{\ti ***}}
\journalid{337}{}
\articleid{11}{14}

\lefthead{Kalkofen}
\righthead{The Case Against Cold, Dark Chromospheres}

	\def\Kv{K$_{\rm 2v}$}
	\def\Hv{H$_{\rm 2v}$}
	\def\sub#1{\hbox{$_{\rm#1}$}}
	\def\ssub#1{\hskip-1pt\hbox{$_{\rm#1}$}}
	\def\arcsec{\hbox{$^{\prime\prime}$}}
	\def\msk{\hskip-1pt}
	\def\gapp{\mathbin{\raise1pt\hbox{$>$}\hskip-		7.5pt\lower3pt\hbox{$\sim$}}}
	\def\lapp{\mathbin{\raise1pt\hbox{$<$}\hskip-		7.5pt\lower3pt\hbox{$\sim$}}}

\begin{document}

\title{The Case Against Cold, Dark Chromospheres}

\author{Wolfgang Kalkofen}
\affil{Harvard-Smithsonian Center for Astrophysics, Cambridge, MA, USA}


\begin{abstract}
Is the solar chromosphere always hot, with relatively small temperature variations ($\delta T/T\sim0.1$); or is it cold most of the time, with temperature fluctuations that reach $\delta T/T\sim 10$ at the top of the chromosphere?
Or, equivalently: Is the chromosphere heated continually, or only for a few seconds once every three minutes?
Two types of empirical model, one essentially time independent and always hot, the other highly time dependent and mostly cold, come to fundamentally different conclusions.
This paper analyzes the time-dependent model of the quiet, nonmagnetic chromosphere by Carlsson \& Stein (1994: CS94) and shows that it predicts deep absorption lines, none of which is observed; intensity fluctuations in the Lyman continuum that are much larger than observed; and time-averaged emission that falls far short of the observed emission.
The paper concludes that the solar chromosphere, while time dependent, is never cold and dark. The same conclusion applies for stellar chromospheres.

A complete, time-dependent model of the nonmagnetic chromosphere must describe two phenomena: (1) dynamics, like that modeled by CS94 for chromospheric bright points but corrected for the geometrical properties of shocks propagating in an upward-expanding channel; and (2) the energetically more important general, sustained heating of the chromosphere, as described by current time-independent empirical models, but modified in the upper photosphere for the formation of molecular absorption lines of CO in a dynamical medium.
This model is always hot and, except for absorption features caused by departures from local thermodynamic equilibrium, shows chromospheric lines only in emission. 

\end{abstract}

\section{Introduction}

Empirical models of the temperature structure of the solar chromosphere have traditionally aimed at reproducing the emergent spectrum. Such models, from the Bilderberg model (Gingerich \& de Jager 1968) to the Harvard-Smithsonian Reference Atmosphere (Gingerich et al.\ 1971) and culminating in the models by Vernazza, Avrett \& Loeser (1981: hereafter VAL81), with improvements in the layers of formation of the emission features in the calcium H line by Avrett (1985), and an extension to the solar transition region by Fontenla, Avrett \& Loeser (1993: hereafter FAL93), have in common a monotonic temperature rise in the outward direction. Inhomogeneous brightness components, such as quiet cell, magnetic network, and active regions, are represented by separate spatial models. 
Intensity variations with time, which can amount to an order of magnitude in the EUV, are accounted for by temperature variations of a few hundred degrees.

A new kind of model, by Carlsson \& Stein (1994: hereafter CS94), simulates the dynamics observed by Lites, Rutten \& Kalkofen (1993: hereafter LRK93) at several (Ca II) \Hv\ bright point positions in the quiet, nonmagnetic chromosphere. 
While the model is successful in matching observed velocity shifts in the H line, it predicts an emergent \Hv\ intensity at maximal brightness that is much higher than observed, a line core in the H line at maximal redshift that is much darker than observed, 
and a time-average emission from all heights in the chromosphere that is much lower than observed.

In addition to reproducing the dynamics, the simulations produce a time-dependent temperature structure that is fundamentally different from the temperature structure of the earlier models. Except for the shock wave that causes the bright points, the CS94 model has a temperature that decreases in the outward direction and is much colder than their starting model or a model by Kurucz (1996), both of which are in radiative equilibrium. The temperature fluctuations in this dynamical model are very large, at times exceeding a factor of ten at the top of the chromosphere. A time average of the temperature structure for an interval shorter than the wave period would show the intermittence of the heating and would not resemble the VAL81 models.

A previous paper (Kalkofen, Ulmschneider \& Avrett 1999) discussed the likely cause and remedy of the energy deficit of the CS94 model. 
The present paper emphasizes the observational evidence against a chromosphere that is heated only intermittently.
Section 2 presents the temperature structure of the VAL81 model and shows that the likely heating mechanism of the solar chromosphere is continuous shock dissipation; Section 3 describes the dynamical simulations of CS94 that provide intermittent shock heating; Section 4 predicts features of the emergent line radiation; Section 5 estimates fluctuations of the intensity in the Lyman continuum; Section 6 considers constraints on models from observations of infrared lines of the CO molecule; and Section 7 discusses various time averages of the time-dependent temperature. In Section 8 I draw conclusions.

\section{Shock Heating of the Quiet Solar Chromosphere}

The several empirical models by VAL81 and FAL93 represent different brightness components of the solar chromosphere. Near the base of the chromosphere, at a height of 0.5 Mm, the spatial resolution of the Skylab observations is 5\arcsec$\times$5\arcsec, and in
the middle chromosphere at about 1 Mm, where the emission in the resonance lines of Ca II arises, the spatial resolution of the H line observations by Cram \& Dam\'e (1983: hereafter CD83) is 500 km (0.$\hskip-3pt^{\prime\prime}7$) and the cadence is 10 s. Both the Skylab and the H line observations figure prominently in the VAL81 models (cf.\ Avrett 1985).
The component models A through F all have typical temperatures near 6,000 K in the middle chromosphere. At a height of 1 Mm, models A and F differ from model C, the average model, by less than 300 K, and the rise to the fully ionized chromosphere occurs in all models within a band of 100 km near the height of 2.2 Mm.

The chromosphere of the quiet Sun shows large variations in the intensity of the emergent radiation field. The corresponding temperature fluctuations are bracketed by models A and F. Although higher temperatures than implied by model F and cooler temperatures than implied by model A occur, these extreme conditions cannot play a significant role in the energy budget of the chromosphere. For the coolest conditions, for example, CD83 remark that they give a darker H line profile than model A does. But even the lowest 10\% of their observations show relatively high emission at line center, H$_3$, which implies that cold conditions can occur only for much less than 10\% of the time or that the temperature drop below that of model A is minor.

Andersen \& Athay (1989) analyzed the average chromospheric model, VAL81 model C. Instead of pursuing the usual aim in theoretical modeling, which had been to reproduce the height-dependent cooling function of the model (Stein 1985), they started from the empirical temperature structure and included all important opacity sources, among them lines of Fe II and other metals, which doubled the opacity used by VAL81. The result of this investigation was the conclusion that the radiative emission rate in the chromosphere is a linear function of mass density, except in the layers near the base of the chromosphere, where the emission increases sharply from its low value in the upper photosphere to its final value in the chromosphere.

Heat conduction is negligible in the layers under discussion, $0.5{\rm Mm}<z<2{\rm Mm}$. Therefore, the dissipation mechanism(s) heating the medium must deliver the energy powering the radiation at the point of emission.

Consider the properties of plane acoustic waves. The energy flux, $F\sub{wave}$, is 
\begin{equation}
	F\ssub{wave}= c\sub{s} \rho v^2,
\end{equation}
where $c\sub{s}$ is the speed of sound,  $\rho$ the mass density and $v$ the velocity fluctuation of the gas. Since the sound speed depends only weakly on temperature (as $\sqrt T$) and the temperature varies only weakly in the chromospheric layers above 1 Mm we may assume that $c\sub{s}$ is constant. We are then concerned with the dependence of $F\ssub{wave}$ on $\rho$ and $v$ only.

In the low-amplitude limit, $v\ll c\sub{s}$,
the energy flux in the wave is constant, hence the velocity grows exponentially with height to compensate for the exponential decay of $\rho$, i.e.,
\begin{equation}
	v\propto \rho^{-1/2} \propto {\rm e}^{z/2\cal H},\quad F\ssub{wave} = {\rm const},
\end{equation}
where $z$ is height above $\tau\sub{5000}=1$ and $\cal H$ is the density scale height of the gas --- for convenience we consider isothermal conditions and one-dimensional wave propagation.

Because of the exponential increase (eqn.\ 2) with height, the velocity grows into the nonlinear regime ($v\gapp c\sub{s}$) until dissipation in shocks limits further growth.
Then the velocity remains approximately constant (Ulmschneider 1991), and the energy flux decays with height like the mass density, i.e.,
\begin{equation}
	F\ssub{wave} \propto \rho, \quad v= \alpha \times c\sub{s}, \quad \alpha \approx {\rm const},
\end{equation}
and the rate at which shocks dissipate energy is proportional to $\rho$,
\begin{equation}
	{{{\rm d}F\ssub{wave}}\over{{\rm d}z}} \propto \rho.
\end{equation}
Thus, heating and cooling rates have the same, linear, density dependence.
Note that this conclusion applies to the instantaneous heating and cooling rates.

The energy flux generated in the convection zone by the Lighthill mechanism (Stein 1967, 1968; Musielak et al.\ 1994) is large enough to cover the radiation losses of the chromosphere. Therefore, acoustic waves are able to  heat the chromosphere. This has been demonstrated with monochromatic acoustic waves (e.g.\ Schmitz, Ulmschneider \& Kalkofen 1985) for a wave period of 30 s; we expect that a continuous wave spectrum achieves the same result. A time average of such models shows a continuous outward temperature rise, resembling that of the VAL81 models.
For a continuous acoustic wave spectrum we expect the instantaneous models to resemble the VAL81 models but also to show the temperature fluctuations inherent in the nonlinear process of shock heating.

Any heating mechanism other than continuous shock dissipation would have to meet the requirement demanded by the VAL81 models, that the dissipation rate is a linear function of mass density. Our result (eqn.\ 4) therefore provides theoretical support for the empirical models built on observations of the emergent chromospheric radiation. Thus, one can argue, the empirical VAL81 models correctly describe the temperature structure of the quiet solar chromosphere. These models allow temperature fluctuations of relatively low amplitude due to sustained, continual shock heating, but they do not describe the huge temperature variations that occur in the CS94 simulations of chromospheric bright points.

\section{The Temperature from Simulations of the Dynamics}

Observations of calcium bright points in the quiet nonmagnetic chromosphere suggest a causal link between velocity shifts of  the Ca II H line in the chromosphere and Doppler motions of an Fe I line in the photosphere immediately below. CS94 simulated this dynamics by taking the velocity fluctuations of the iron line from a one-hour observing run of LRK93 as a condition at the lower boundary in their radiation hydrodynamics code, and determined the subsequent velocity shifts in the H line. Except for a brief time interval of the order of the three-minute wave period when the waves enter an undisturbed, initial atmosphere, the simulations reproduce the characteristic features of H line behavior, such as the occurrence of maximal brightening of the blue emission peak, \Hv, simultaneously with maximal redshift of the line center, H$_3$. 
Discrepancies of the simulations with the observations of LRK93 in the timing of the velocity shifts in the line profile, although they amount to a substantial fraction (1/3) of the sound travel time, are of minor significance, allowing the conclusion that the underlying physics of the model is correct; but major discrepancies in the \Hv\ and H$_3$ intensities suggest that the underlying temperature structure and the geometry of wave propagation are different from those of the Sun.

The dynamical simulations also yield a time-dependent temperature structure (Figure 1), with extraordinary properties, from which CS94 and Rutten (1994) concluded that the solar chromosphere does not have a steady temperature rise. This was further emphasized in a paper by Carlsson \& Stein (1995: hereafter CS95) titled ``Does a Nonmagnetic Solar Chromosphere Exist?''  followed by a paper (Carlsson \& Stein 1998) titled ``The New Chromosphere.''  The conclusions from this modeling were extended to cool stars in general by Schrijver (2000) who stated that ``there is no such thing as a steady temperature minimum over the top of their \hbox{photospheres.\msk\msk''}
 

The salient features of the dynamical temperature structure, shown in Figure 1, and of a snapshot, shown in Figure 2,
are a lower envelope for the cool background, with the temperature decreasing monotonically with height; an upper envelope for the temperature excursions due to the upward-propagating shock wave, with the amplitude of the temperature fluctuations generally increasing with height; a time-average representing the straight time average throughout the atmosphere, with the temperature  decreasing monotonically with height; and two monotonically increasing temperature distributions, one corresponding to the time-average emission of the dynamical CS94 model (the Semi-empirical curve in Figure 1) and the other, to the time-average emission of the Sun (FALA).

The temperature distributions in Figure 1 allow us to infer a cooling time for the medium and to construct a two-component model, consisting of a cool background and a thin, hot slab moving upward over it (see Figure 2, from CS95). 
We estimate the cooling time from the fraction of the time the atmosphere is in the high state by noting that the maximal temperature at the top of the atmosphere is $T\sub{max}=25,000$ K, the minimal temperature is $T\sub{min}=2,500$ K (see CS95) and the average is $T\sub{av}=3,800$ K. The temperature differences ($T\sub{max}-T\sub{av}$) and ($T\sub{av}-T\sub{min}$) are in the ratio of 20:1, implying, grosso modo, that the times spent by the atmosphere in the high and low states are in the ratio of 1:20 and that the cooling time must therefore be of the order of one twentieth of the wave period.
The temperature behavior behind the shock that is suggested by Figure 2 is a drop that is linear in time and begins as soon as the temperature reaches its maximal value; it gives a fraction of 88\% for the time interval in the cold state and 6\% for the time interval during which the temperature is above the halfway point between maximum and minimum. From the wave period of three minutes and the sound speed of 7 ${\rm km{\hskip1pt}s}^{-1}$, which is the wave speed in the linear limit, we estimate a cooling time of 10 s and a thickness of the slab at the halfway height of 70 km.

The actual simulations give the temperature as a function of height at a fixed time, and as a function of time at a fixed height:
Figure 2, from Fig.\ 1 of CS95, shows a snapshot of a simulation with the temperature profile as a function of height when the shock is at 1.4 Mm. The thickness of the hot slab, defined as its width at half maximum, is approximately 80 km.
And in Figure 3, from Fig.\ 4 of CS94, the temperature variation as a function of time at unit optical depth at the Lyman absorption edge shows that the temperature peaks have a width of about 10\% of the wave period. The cooling time in the layer of formation of the Lyman continuum is therefore of the order of 20~s.

We note that the degree of ionization of hydrogen during the oscillations remains near $10^{-1}$ (see Fig.\ 2 of Carlsson \& Stein 1999), implying a recombination time to the hydrogen ground state that is long compared to the wave period. The cooling time we have estimated therefore describes only the behavior of the kinetic temperature. Although the atmosphere between passages of shocks may achieve approximate balance between radiative heating and cooling, the state of true radiative equilibrium is not reached because of the long radiative relaxation time of hydrogen.



All estimates of the cooling time concur in implying a thin hot layer representing the shock moving outward over a background in which the temperature decreases monotonically with height.
We may therefore represent the time-dependent kinetic temperature of the chromosphere schematically by a two-component model, consisting of a time-independent cool background and a time-dependent hot slab with a thickness of about 10\% of the thickness of the chromosphere traveling upward over the background,
\begin{equation}
T\sub{CS94}(t,z) \approx T_{0}(z) + \delta T(t,z),
\end{equation}
where $T_{0}(z)$ is described by the lower envelope in Figure 1 and the maximal excursions $\delta T(t,z)$ by the upper envelope.

\section{Instantaneous Emergent Radiation in Lines}

The wave period observed in the calcium bright points is approximately equal to three minutes, and the wave travel time through the chromosphere (the region extending from 0.5 Mm to 2 Mm) is also approximately equal to three minutes. Thus, at any instant of time, there is typically only one strong shock traveling through the chromosphere.
Furthermore, because of spatial intermittence (Lites, Rutten \& Thomas 1994), larger areas in the chromosphere may show no high-amplitude three-minute oscillations, and because of temporal intermittence (von Uexk\"ull \& Kneer 1995), trains of high-amplitude oscillations may be separated by time intervals of a few times the wave period. When strong shocks are absent, only the naked background is found, which in the CS94 model has a monotonically declining temperature, i.e., $T\sub{CS94}(t,z) \approx T\sub{0}(z)$.

An interesting quantity to estimate and compare with observations is the residual intensity in the core of the H line, i.e., the ratio of the emergent intensities in the line and the neighboring continuum.
For a typical scattering line the emergent intensity is approximately equal to the line source function. Taking 3,000 K for the temperature in the upper layers (see Figure 1), assumed to be isothermal and static,  the source function in CRD at the top is given by $S(0)=\sqrt{\epsilon}\ B_\nu(3,000 {\rm K})$, where $\epsilon$ is the scattering parameter, with a value of $\epsilon=0.02$ in the FAL93 model (Avrett, private communication), and $B_\nu(T)$ is the Planck function. The assumption that the medium is approximately isothermal must hold for a layer with a thickness equal to the thermalization length, which for a Doppler-broadened line in a static medium is equal to $\tau\ssub{relax}=1/\epsilon$.
Since the line center optical depth at the height where the line source function has its maximum, near 1 Mm, is $\tau_3=500$ and thus ten times the thermalization depth, the emergent intensity should be adequately represented by the isothermal approximation when a shock is in the low chromosphere or is absent altogether.

We calculate the intensity of the continuum radiation near the H line as the Planck function for the observed brightness temperature of  4,677 K near the H and K lines (Avrett, priv.\ comm.), which gives a residual intensity at the center of the H line of $2\times10^{-3}$.
Note that this estimate is an upper limit because of the assumption that the layer above 1 Mm is isothermal at $T=3,000$ K and does not drop to 2,000 K at the top.
We should observe this or a lower value of the residual intensity at line center for any position or time without large-amplitude oscillations, as well as at positions of bright points when the shock is in the low chromosphere, below the peak of the \Hv\ emission; this occurs  for about one third of the wave period at bright point locations during active oscillations, and generally for about 50\% of bright point locations (see Carlsson, Judge \& Wilhelm 1997). 

The prediction for the residual intensity at H$_3$ is not borne out by the observations of CD83, shown in Figure 4. The profiles for  positions of the shock in the lowest layers of the chromosphere, or for locations and times without large-amplitude oscillations, are found in the deciles for the lowest values of  H index and residual intensity.
At least six of the ten bins should be dominated by profiles with  line radiation from the cold upper layers. But instead of the predicted value of $2\times10^{-3}$, Figure 4 shows .03 for the lowest decile  and .06 for the highest.
Our prediction agrees qualitatively with the actual simulations (Carlsson \& Stein 1997, 1998), which show a very deep and dark line core, much darker than the observations of LRK93, while at the same time exhibiting an \Hv\ intensity at maximum that is much brighter than the observations.
But neither our prediction from the CS94 simulations nor the simulations themselves agree with the time-resolved observations of the H line by LRK93 (see Figure 5 of Carlsson \& Stein 1998) or Kariyappa, Sivaraman \& Anadaram (1994), or of the K line by Liu (1974, shown in Fig.\ 5).


The spatial resolution of the CD83 observations of 500 km relative to the size of strong bright points of 1,800 km to 2,700 km (CD83, Plate 10) and the cadence of 10 s relative to the wave period of 200 s are sufficient to observe the cold, dark top of the CS94 model. The failure to match the observations allows only one conclusion: The chromosphere is not predominantly cold.

EUV observations with SUMER by Carlsson et al.\ (1997) lead to the same conclusion. The conditions for the formation of the resonance lines of neutral and singly ionized C, N and O are similar to those of the resonance lines of ionized calcium. For the carbon line at 1657 \AA, for example, the line center optical depth at the source function maximum, near a height of 1 Mm, is equal to 200 in the FAL93 model (Avrett, private communication). 

The slit in the SUMER observations measured 1\arcsec$\times$120\arcsec, with sampling every arcsecond along the slit. Although the slit cut through several network arches, most of the positions were located in the nonmagnetic internetwork regions, for a total of about $10^2$ positions. The observing run covered four hours, but each of ten lines was measured for only one hour, corresponding to about twenty times the wave period. These observations therefore constitute $2 \times 10^4$ complete line segments covering a full wave period each. Fifty percent of the chromosphere exhibited the three-minute oscillations. Thus, $10^4$ segments should show the complete passage of a shock wave through the chromosphere, and $10^4$ segments, corresponding to locations without large-amplitude oscillations, should originate in a chromosphere without a shock. Consequently, half the segments should show the transition from emission to absorption lines, and the remaining segments should show only deep absorption.
Carlsson et al.\ (1997) found only emission lines.

While it would be conceivable that emission from a hot canopy might fill in some of the absorption predicted for the CS94 model, it is inconceivable that it would fill in every single one of the $2\times10^4$ segments and, furthermore, turn deep absorption lines completely into emission lines, even if the canopy had a filling factor of unity (see Jones 1985). The requirement to turn absorption into emission allows no exception since, as the authors note, ``all chromospheric lines show emission above the continuum everywhere, all of the time.'' The firm conclusion that the chromosphere is never cold is inescapable.

Finally, the K line observed by Liu (1974), in Figure 5, shows the emergent intensity at three different phases of a wave. In the relaxed phase, denoted by $t=0$, the chromosphere is without a wave and the profile is symmetric. Since the intensity represents the mapping of the temperature structure from depth to wavelength, its increase from the K$_1$ minima toward line center results from the chromospheric temperature increase outward while the central absorption results from scattering, which decouples the source function from the Planck function. The traditional interpretation of this profile is that the temperature rises in the outward direction.
The observed line profile implies a well-defined temperature minimum at the top of the solar photosphere and thus contradicts the claim to the contrary by CS94 and Schrijver (2000).


\section{Intensity Fluctuations in the Lyman Continuum}

The intermittent heating seen in Figures 2 and 3 and described by equation (5) leaves its imprint also on the emergent intensity in the Lyman continuum. Although the CS94 simulations compute the Lyman radiation only at the absorption edge, their paper provides sufficient information to determine the emergent Lyman continuum spectrum also at other wavelengths. The model predictions can then be compared with Skylab observations.

Consider the second time interval in Figure 3, which covers a complete wave period. Although the kinetic temperature reaches $10^4$ K and then drops to 2,000 K, the brightness temperature, which describes the emergent intensity at the Lyman edge, varies by only a relatively small amount about the average of 6,250 K, between 6,600 K and 5,750 K. But in the intensity this temperature variation is magnified and amounts to an intensity ratio of a factor of 34, which is about four times the value seen in the Skylab observations that are represented in the empirical VAL81 models A to F (see Fig.\ 9 of VAL81).

The emergent intensity at shorter wavelengths can be calculated from the intensity at the absorption edge and the kinetic temperature of the electrons. Since the time for thermal relaxation of the electrons by Coulomb collisions is very short --- it is measured in microseconds (Spitzer 1956) --- the electrons always satisfy a Maxwellian velocity distribution at the local kinetic temperature, and the emitted Lyman photons always follow a Planckian frequency distribution. The shape of the emergent intensity is therefore known. It is given by the Planck function at the kinetic temperature approximately at unit monochromatic optical depth. Since the opacity in the Lyman continuum is a weak function of wavelength ($\sim \lambda^3$), unit optical depth at $\lambda$ is close to unit optical depth at 912 \AA\ if $\lambda$ is close to the absorption edge. For our estimate we ignore the difference in depth and take the temperature at $\tau\sub{Lc}=1$ as a close approximation to the temperature at $\tau_\lambda=1$. But it should be noted that while the temperatures at the two depths may be close, the source functions need not be so close. Since the electron kinetic energy at a temperature of 2,000 K is only of the order of a percent of the hydrogen ionization energy of 13.6 eV, a small temperature difference is magnified by the exponential dependence of the Planck function on temperature and wavelength. But for a rough estimate of the intensity fluctuation we may ignore this effect.

Early in the second time interval in Figure 3, at time $t_1$, the curves for kinetic temperature, brightness temperature, the source function at $\tau\sub{Lc}=1$, and the time-average brightness temperature cross at a temperature of $T\sub{kin,1}=6,250$ K. The emergent intensity at wavelength $\lambda$ is therefore given by the Planck function at $T\sub{kin,1}$,
\begin{equation}
	I_\lambda(t_1)=B_\lambda(T\sub{kin,1}).
\end{equation}
Later in the time interval, at time $t_2$, when the brightness and kinetic temperatures are $T\sub{br,2}$ and $T\sub{kin,2}$, the emergent intensity at wavelength $\lambda$ is given by
\begin{equation}
	I_\lambda(t_2)=B_\lambda(T\sub{kin,2})\times B_{912}(T\sub{br,2})/B_{912}(T\sub{kin,2}).
\end{equation}
For the time $t_2$ we choose the instant when the kinetic temperature is $T\sub{kin,2}=2,200$ K and the brightness temperature is $T\sub{br,2}=6,000$ K. Since the source function at $t_2$ is lower than the emergent intensity, the actual ratio of intensities at $t_1$ and $t_2$ is larger than our estimate.

For the kinetic temperatures of 6,250 K and 2,200 K we predict intensity ratios $I(t_1)/I(t_2)$ of 2.9 at 912 \AA, $1.9\times10^3$ at 800 \AA, and $1.4\times10^5$ at 740 \AA. The observed ratios at these wavelengths in the Skylab data shown in Fig.\ 9 of VAL81 are less than a factor of ten. Our estimates are very much larger than any observed. We note that the temperature variation in the four wave segments shown in Figure 3 is larger than the variation used in the estimate. The intensity fluctuations of the model therefore exceed the prediction. 

The Skylab observations have a spatial resolution of 5\arcsec$\times$5\arcsec\ and a cadence of either one minute or 41 ms. Spatial or temporal averaging can therefore not account for the difference between observed and predicted intensity fluctuations in the Lyman continuum. We are led to conclude that the solar chromosphere suffers much smaller temperature variations than the model and is not heated only intermittently by large-amplitude waves.

\section{Constraints on Models from CO Lines}

Lines of the carbon monoxide molecule, in emission off the limb and in absorption on the disk, have never been successfully explained in terms of an accepted empirical model, except by invoking a separate model, such as the COmosphere (Wiedemann et al.\ 1994). It is interesting to ask whether the dynamical model of CS94 succeeds where static models have failed.

Observations of the vibration-rotation bands of CO by Noyes \& Hall (1972) and Ayres (1981) indicate temperatures for line formation as low as 4,100 K at disk center and near 3,800 K close to the limb, and Ayres \& Testermann (1981) noted that the CO brightness temperatures are incompatible with a chromospheric temperature rise as in the FAL93 models (see also Avrett 1995). Instead, these observations are believed to require a temperature that is monotonically decreasing in the atmospheric layers where current empirical models place the chromospheric temperature increase.
For the line formation problem in CO, the CS94 model appears to offer a solution since the background atmosphere has a monotonically declining temperature profile.

The observational data to be fitted by a model are (1) the formation height of the CO absorption (or emission) lines, (2) the temperature of the gas at that height, and (3) the temperature fluctuations observed in the line intensity.

Observations of emission in CO off the limb by Ayres (1998) suggest that molecular gas is present up to a height of 900 km, and perhaps up to 1 Mm or more, with a temperature of 3,500 K. While this temperature agrees at the quoted  height with that of the background model in the CS94 simulations, the temperature at the shock front reaches 6,300 K at 900 km and $10^4$ K at 1.1 Mm, virtually guaranteeing the complete destruction of CO molecules by the shocks responsible for chromospheric bright points. With an association time of hours (Avrett et al.\ 1996), CO could not exist at such heights. 
The interpretation of the limb observations in CO lines is thus incompatible with the temperature structure of the CS94 model.
Modeling by Uitenbroek (2000) of CO lines on the basis of snapshots of the dynamical model gives an amplitude of intensity variations that is higher than observed by a factor of 2.5 and therefore leads to the same conclusion.

The temperature fluctuations of $\pm300$ K observed by Ayres (1998) in CO emission lines are broadly consistent with observations of CO absorption lines by  Uitenbroek \& Noyes (1994) and Uitenbroek, Noyes \& Rabin (1994) who note that while there are cold elements that are as much as 200 K colder than the average, the bulk of the temperature variation remains within $\pm100$K. The height at which a peak-to-peak temperature variation of 400 K is found in the CS94 model is 500 km. But the time-average temperature at that height is 4,800 K (see Figure 1), which is much higher than the values suggested by the observations.

Thus, none of the parameter values obtained from analyses of CO observations on the basis of static atmospheres fits into the dynamical CS94 model.
The remedy of that model proposed by Kalkofen et al.\ (1999) cannot help since its effect is to raise the background temperature from the low values of the cool CS94 background model to the higher values of the FAL93 model. 

It is unlikely that calcium bright points and molecular CO absorption lines occur in different regions on the Sun --- the latter, for example, in the isolated, much colder structures suggested by the so-called COmosphere or by the two-component model of Ayres, Testermann \& Brault (1986), which overestimates the UV continuum flux at 1,400 \AA\ by a factor of 20 (see Avrett 1995). It is much more likely that CO absorption lines and bright points occur in the same regions, but at different heights. The reason is that bright point oscillations are pervasive, with a filling factor of 50\% (Carlsson et al.\ 1997), and the CO-absorbing regions are even more pervasive, with a filling factor of 50\% to 85\% (Solanki, Livingston \& Ayres 1994).

Given these considerations, a plausible scenario for CO lines is that they are formed in a dynamical medium in the layers of the upper photosphere. This dynamical atmosphere combines features of both the time-dependent CS94 model and the time-independent semiempirical FAL93 model A, the former describing the temperature fluctuations and the latter the low temperatures in the temperature minimum region, which may be caused by cooling due to granular motion (Stein \& Nordlund 1989). In these layers, important constraints on the temperature structure of the VAL81 and FAL93 models come from observations of the calcium lines (CD83, Ayres \& Linsky 1978) and EUV continua (VAL81) but not from CO lines. A lower empirical temperature in the temperature minimum region might be compatible with the UV data as well as with CO line formation in a dynamical atmosphere.
The minimum temperature of 4,240 K of FAL93 model A with superposed velocity and temperature fluctuations, the latter with an amplitude of 200 K, might be consistent with CO observations at disk center (Avrett, priv.\ comm.), but the problem of the low brightness temperatures seen at the limb remains.

\section{Time-Average Models}

The various FAL93 models differ from one another by a few hundred degrees in the ``chromospheric plateau'' region, i.e., at heights between 1 Mm and 2 Mm and temperatures between 6,000 K and 7,000 K, where they are based mainly on the CD83 data.
Since the H line profiles are placed into ten ordered bins, the highest and lowest observed intensities are represented only in an average way in the FAL93 models. 
Very high and very low intensities do not play a significant role in the energy budget of the chromosphere, however. Otherwise the top and bottom bins (see Figure 4) would differ significantly from the others. Thus, apart from the intermittent heating due to the passage of the strong shock responsible for bright points and apart from the relatively small temperature fluctuations inherent in the sustained shock heating of the atmosphere, the FAL93 models as a group correspond to time averages over the 10 s cadence of the CD83 observations.

CS94 show two kinds of average temperature (see Figure 1): 1.\ the straight time average of the fluctuating temperature, and 2.\ the temperature of a static model that matches the time-averaged emission from the time-dependent model. The latter model is thus analogous to the (semiempirical) FAL93 models, except that whereas the CS94 model is designed to match the time-averaged radiation from the dynamical model, the FAL93 models are designed to match the time-averaged radiation from the Sun.

If the time averages of the CS94 model, either the straight average or the average emission (``Semi-empirical'' in Figure 1), were also taken over 10 s they would reflect the intermittent heating pattern and would therefore depend on the phase of the shock. 
If they were taken over a full wave period they would show the considerable variation of the peak intensity in the calcium emission peaks (Figs.\ 2 and 3 of von Uexk\"ull \& Kneer 1995). The FAL93 models would not have the same variation since the solar chromosphere always shows background emission in the calcium lines. 
It seems likely that the time averages in Fig.\ 5 of CS94  (Fig.\ 1 in this paper) were taken over the observation time of one hour, perhaps excluding the initial, transient behavior. 

Another noteworthy feature is that the averages in the CS94 model are taken over at most four very strong bright points, and possibly over only the most luminous bright point in  Fig.\ 1 of LRK93, which far outshines any network bright point at that observation time and generally matches the brightest network emission in the observing run (Fig.\ 2 of LRK93). For such a bright point, the appropriate comparison model of the FAL93 series would be model C or a hotter model. Instead, CS94 compare their model with the coolest of the VAL81 models, FAL93 model A.
In spite of this bias, the CS94 temperature falls below the FALA temperature by a considerable margin (see Fig.\ 1), implying a considerable deficit in total energy emitted from the chromosphere.

One can estimate the deficit, at a height of 1 Mm for example, from the temperature curves for the semiempirical models in Figure 1. The most important emitters in the chromosphere are the lines of Ca II and Mg II (see VAL81, Fig.\ 49). Therefore, measuring the time-average emission by the Planck functions for the H and K lines of Ca II, which up to that height are tracked by the line source functions, the shortfall of the CS94 model relative to model FALA is a factor of 3, and for the h and k lines of Mg II it is a factor of 5.
The shortfall would be even larger in the more appropriate comparison with a FAL93 model hotter than FAL-A.
This finding agrees qualitatively with the observational result of Hofmann, Steffens \& Deubner (1996), discussed by Kalkofen et al.\ (1999), that the emission  in \Kv\ during the bright point phase is only 9\% of the total \Kv\ emission from the nonmagnetic Sun. But the deficit apparent in Figure 1 is smaller than the factor of 10 given by these K line observations; this supports our supposition that CS94 have modeled exceptionally bright features.

The lower value of the straight time-average temperature of the CS94 model compared with the temperature of the FAL93 model produces a smaller scale height and, consequently, a smaller thickness of the chromosphere. Thus, the upper boundary of the CS94 model is not at 2.2 Mm, as in the FAL93 models, but at the lower height of 1.8 Mm. This difference is a consequence of the difference in temperatures and, hence, scale heights. (Note that the FAL-A model has been redrawn in Fig.\ 5 of CS94 and the location of its top has been lowered from its original height of 2.2 Mm in FAL93 model A down to 1.9 Mm. Note also  that the Semi-empirical curve in Fig. 1 is not defined above 1.8 Mm).

\section{Conclusions}

The modeling of the dynamics of \Hv\ bright points in the nonmagnetic solar chromosphere by CS94 yielded as a by-product a time-dependent temperature that varies between very low and very high values (2,000 K and 25,000 K at a height of 1.8 Mm).
Predictions of radiation from this model are of lines that alternate with the phase of the upward-propagating shock wave between very strong emission lines and very deep absorption lines;
for locations and times without large-amplitude oscillations, which at any instant of time have a spatial filling factor of 50\%, only absorption lines are predicted.
But none of the absorption lines, which should be among the strongest lines in the solar spectrum, has been observed, either from the ground or from space.
For the emergent Lyman continuum radiation, the intensity for the single wavelength point (912 \AA) given by the model agrees with the observations within a factor of four, but the decrease of the intensity with decreasing wavelength is much steeper than observed. 
As a consequence, the intensity fluctuations predicted for the CS94 model are much larger than observed, exceeding a factor of $10^5$ at a wavelength of 740 \AA, where the Skylab observations show a factor of less than ten.

The temperature characterizing the average emission of the CS94 model reflects the intermittence of the heating by a shock wave if the average is determined for a time interval shorter than the wave period. For longer intervals the temperature shows a monotonic rise in the outward direction and therefore agrees qualitatively with the FAL93 models. 
Since the CS94 model is constructed for the brightest features in the LRK93 observations of the quiet, nonmagnetic Sun it corresponds to a FAL93 model that is hotter than their average model, which is FAL-C.
But instead of being hotter, the semiempirical CS94 model is everywhere cooler than even the coolest of the FAL93 models, their model A.
This indicates a significant deficit in the radiated power of the CS94 model relative to the Sun.
We estimate this deficit of the CS94 model on the basis of the H and K lines of Ca II as a factor of 3, and on the basis of the h and k lines of Mg II as a factor of 5 relative to FAL93 model A, and larger factors relative to the Sun. This conclusion is not affected by seeing, scattered light, or a magnetic canopy.

Carlsson \& Stein have modeled only one aspect of chromospheric physics, namely, bright-point dynamics in nonmagnetic regions. Their model contains the intermittent shock heating that is responsible for the dynamics, but not the energetically much more important sustained heating required for most of the radiation emitted by the chromosphere.
While their simulations give a valid physical account of the characteristic velocity signal of bright-point dynamics, the resulting temperature structure does not match the temperature structure of the solar chromosphere.
I conclude that the solar chromosphere is never as cold and dark as they have proposed. 

A complete, time-dependent model of the nonmagnetic solar chromosphere would combine features of the time-independent models of FAL93 and of the time-dependent model of CS94. The main component of this model would describe the sustained heating of the chromosphere and the corresponding emission, but without the contribution of the highly time-dependent emission due to the shock waves that cause the calcium bright points. Thus, the underlying model could be  model A of FAL93, but modified in the temperature minimum region for the time-dependent formation of the vibration-rotation lines of the CO molecule. The dynamics of the combined model would be described by the CS94 simulations, but modified to take account of the geometry of wave propagation for bright points in a stratified medium (Bodo et al.\ 2000). This complete model should rectify two defects of the CS94 model, namely, the deficit of total emission, and the excess of calcium emission at maximal brightening.

\acknowledgements

I thank E.\ H.\ Avrett and P.\ Ulmschneider for discussions and comments on the manuscript, M.\ Carlsson for clarifying statements concerning his model, and the referee for a spirited debate that helped to expand the scope and strengthen the arguments of the paper. I also thank the 
Institut f\"ur Theoretische Astrophysik of the University of Heidelberg for
its hospitality.
Support by NASA and DFG is acknowledged.


\par\eject


\begin{figure}
\plotone{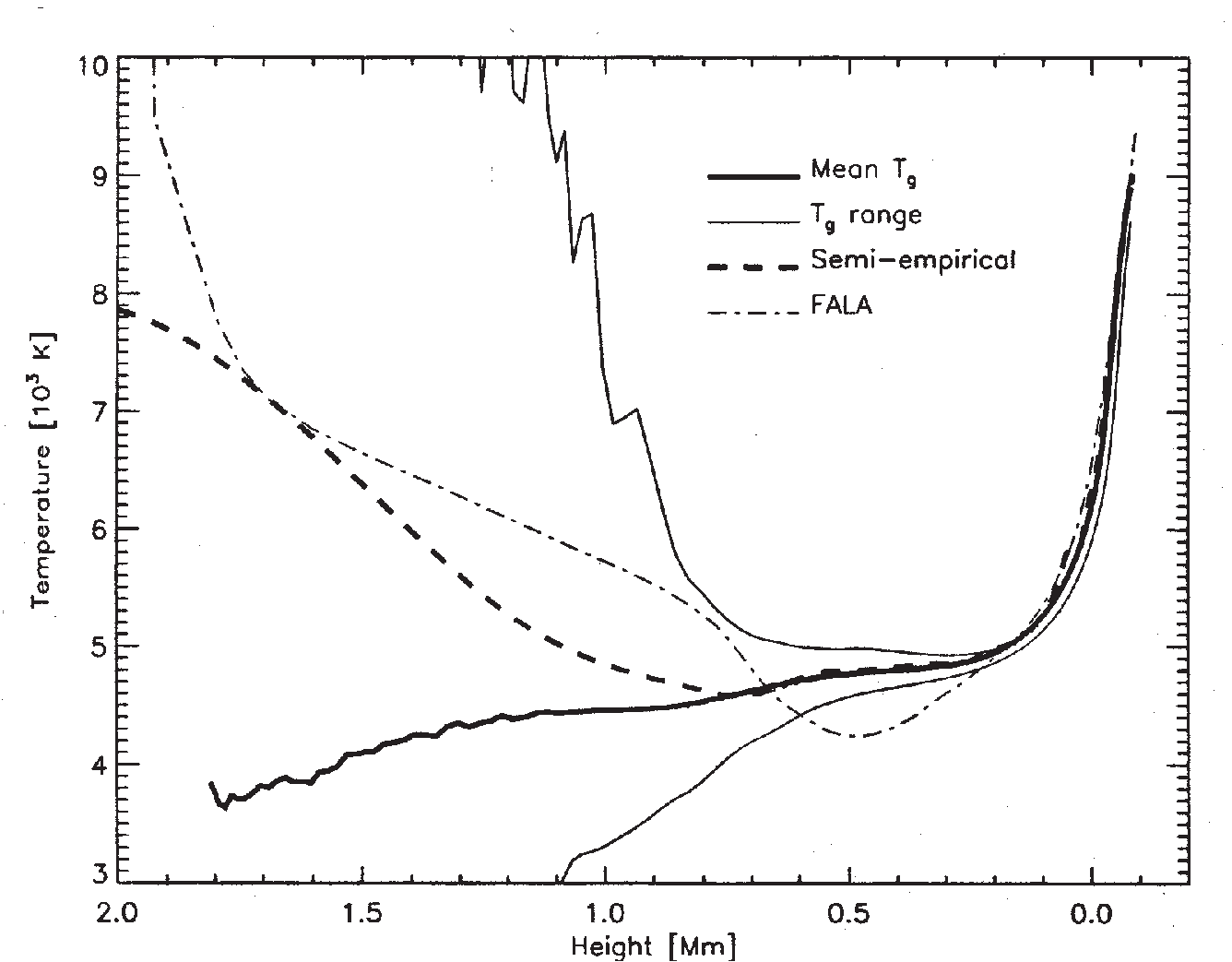}
\caption{Temperatures from simulations by CS94 of the dynamics observed by LRK93: cool background model (lower solid curve); upper temperature excursion of the hot shock (upper solid curve); time average of the temperature fluctuations (thick solid curve); temperature representing the emission averaged over the observation time of one hour, labeled Semi-empirical (heavy dashed curve); and the coolest of the FAL93 models (dash-dot curve)
 ---  from CS94.}
\label{Fig1}
\end{figure}

\begin{figure}
\plotone{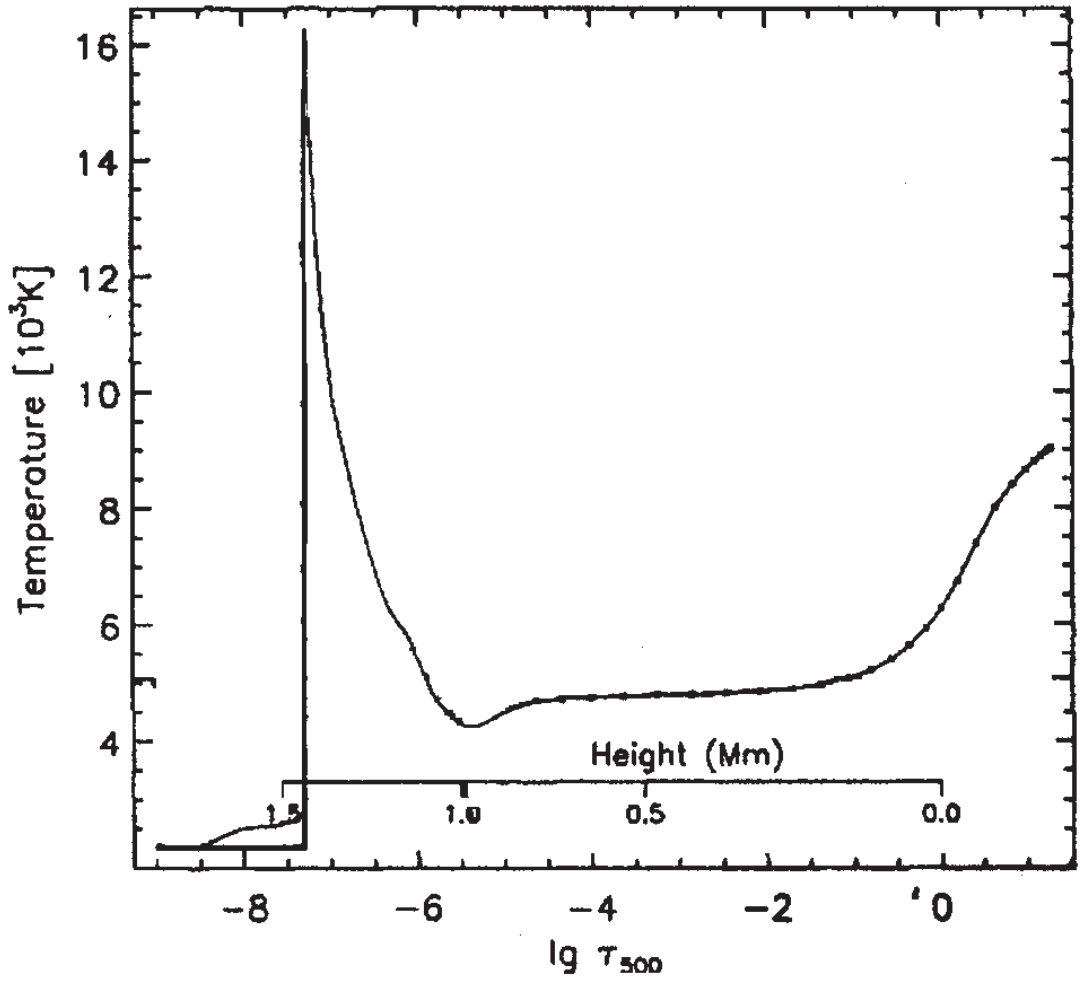}
\caption{Snapshot from a dynamical simulation, showing a single temperature spike at 1.4 Mm. At the top ($z>1.7$ Mm), the temperature drops to 2,000 K ---  from CS95.}
\label{Fig2}
\end{figure}

\begin{figure}
\plotone{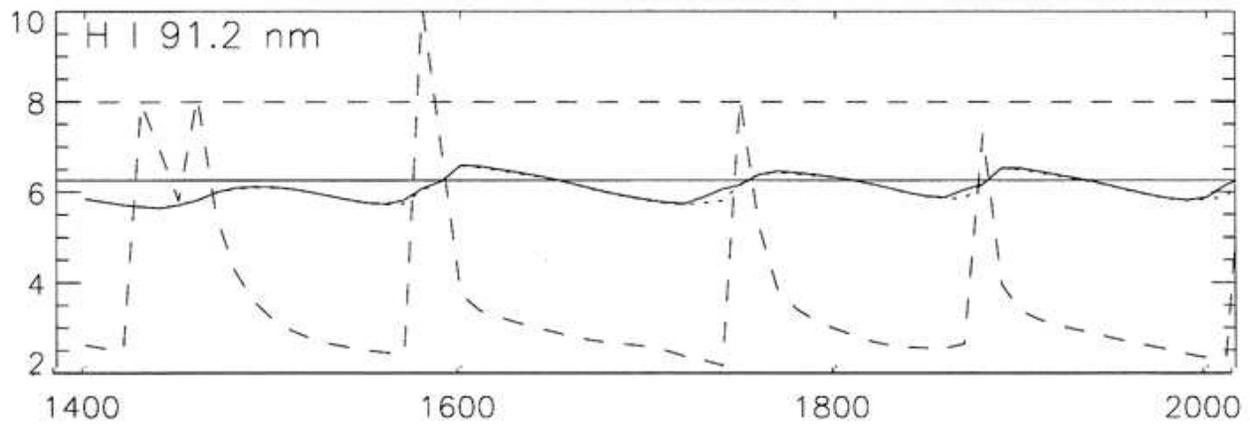}
\caption{Temperatures [in $10^3$ K]  as functions of time [in seconds] at optical depth unity at the Lyman absorption edge: the solid curve is the brightness temperature of the emergent intensity at $\lambda=912$ \AA; the dotted curve is the source function at $\tau\sub{LC}=1$; the dashed curve is the kinetic temperature at $\tau\sub{LC}=1$; the solid horizontal line is the time average of the brightness temperature during the time segment of the figure; and the dashed horizontal line is the temperature corresponding to the time-average of the Planck function --- from CS94.}
\label{Fig3}
\end{figure}

\begin{figure}
\plotone{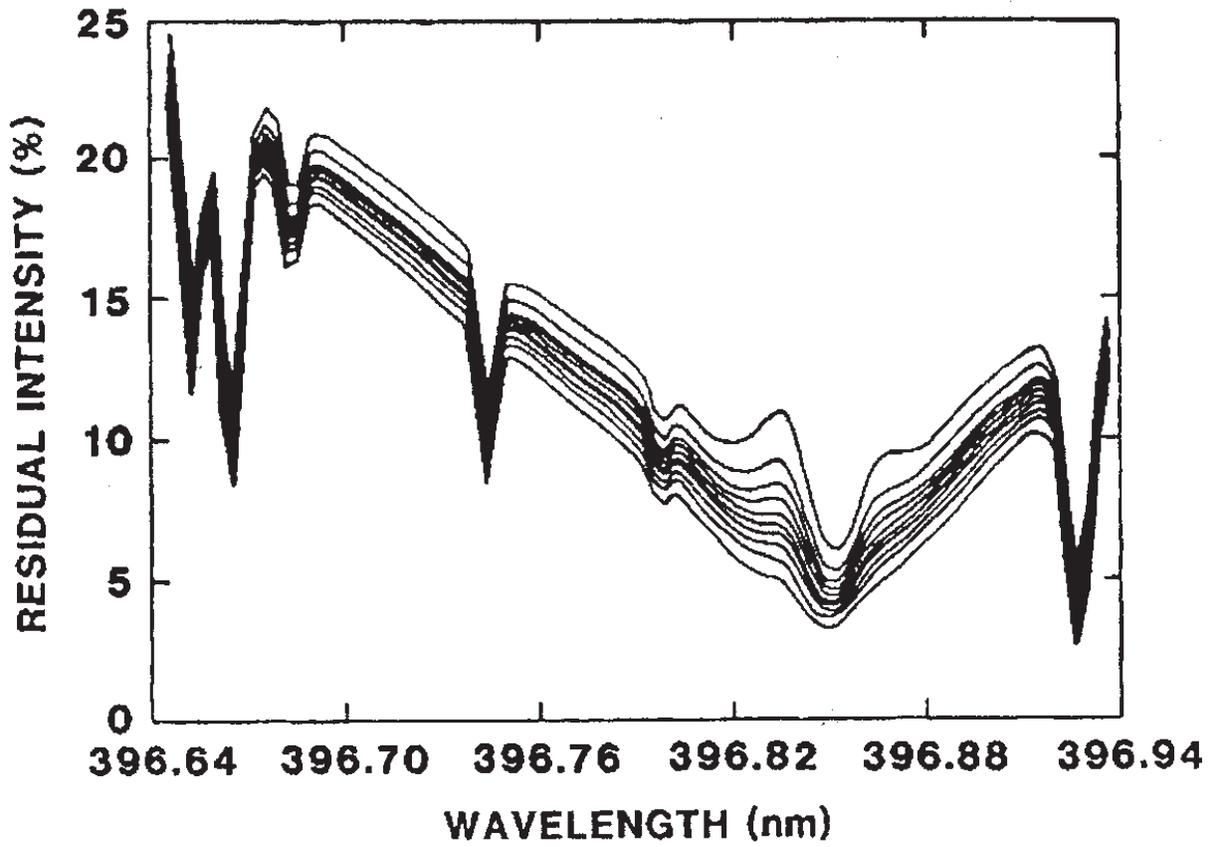}
\caption{Residual intensity in the H line observed by CD83. The profiles are ordered by the H index (emission in a 1 \AA\ band about line center) and placed into ten equal bins. Scattered light (3.8\% of continuum) has been subtracted --- from CD83.}
\label{Fig4}
\end{figure}

\begin{figure}
\plotone{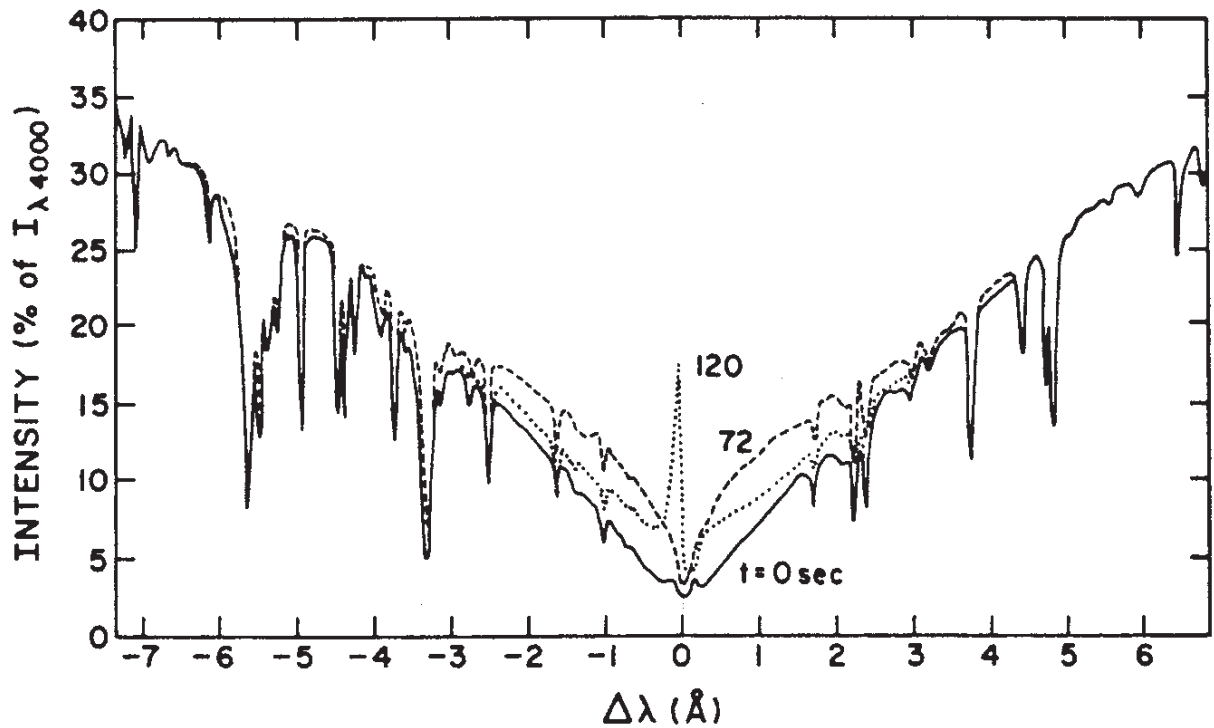}
\caption{The evolution of a \Kv\ bright point at three instants of time. In the relaxed state ($t=0$), the profile is nearly symmetric and shows a chromospheric temperature rise that is not associated with bright point dynamics --- from Liu (1974).}
\label{Fig5}
\end{figure}

\end{document}